\documentclass{article}
\usepackage{spconf,amsmath,graphicx,hyperref}

\usepackage{cite}
\usepackage{amssymb,amsfonts}
\usepackage{algorithmic}
\usepackage{textcomp}
\usepackage{xcolor}
\usepackage{booktabs}
\usepackage{multirow}
\usepackage{colortbl}
\usepackage{caption}
\usepackage{subcaption}
\usepackage{orcidlink}

\title{PerMix: ANOMALOUS SOUND DETECTION BY PREDICTING \\ THE PERTURBATION RATIO OF MIXED SAMPLES}
\name{Jihoon Choi$^1$, Jung-Woo Choi$^{1,2*}$\thanks{*Corresponding Author}}

\address{$^{1}$School of Electrical Engineering, KAIST, Daejeon, South Korea \\
$^{2}$Department of AI Systems, KAIST, Daejeon, South Korea}

\begin{document}
%
\maketitle
\begin{abstract}
In anomalous sound detection based on self-supervised learning, the performance of the detection model largely depends on the selection of the pretext task through which the characteristics of normal data are learned.
A common pretext task is to mix samples from different classes and classify the mixed sample as its target class with an angular margin loss.
However, this supervision becomes mismatched when the target sample is the minority component of the mixture and the margin remains fixed irrespective of the mixing ratio.
To address these issues, we propose a pretext task that directly predicts the perturbation ratio of mixed samples by reformulating the logit so that the angular margin is scaled by the perturbation ratio.
In addition, we adopt a supervised contrastive loss modified for mixed samples, in which positive pairs are formed according to the dominant class.
Experiments show that our approach achieves state-of-the-art results on the DCASE 2020 Task 2 benchmark.
Our code is available at \url{https://github.com/kaistss/PerMix}.

\end{abstract}
\section{Introduction}\label{sec:intro}

Anomalous sound detection (ASD) is a task aimed at identifying sounds that exhibit patterns or features deviating from normal acoustic behavior~\cite{koizumi2020description, dohi2022description}. 
Due to the scarcity and high variability of anomalous samples, self-supervised learning that trains models exclusively on normal data has become the predominant paradigm~\cite{jing2020self, liu2021self}.
In particular, classification-based self-supervised learning has been widely used for ASD, as the representations learned through classification tasks effectively disentangle machine characteristics even in noisy environments~\cite{wilkinghoff2024why}.

Classification-based approaches train the model to classify metadata such as machine IDs or operating conditions, and angular margin losses~\cite{deng2019arcface, wilkinghoff2021sub, choi2024noisy} are commonly employed to enhance intra-class compactness and inter-class separation.
Subsequent studies have added multiple feature extractors to lightweight encoders such as MobileFaceNet~\cite{liu2022anomalous, choi2024noisy, huang2025noise} or fine-tuned large audio models such as BEATs~\cite{chen2022beats, jiang2024anopatch}.
However, enforcing compact class distributions alone can reduce sensitivity to small perturbations near class boundaries, where subtle anomalies may occur.
To address this, mixup~\cite{zhang2018mixup} has been combined with angular margin losses.
NAMix~\cite{choi2024noisy} mixes samples from different classes and trains the model to predict the mixing ratio, applying an asymmetric angular margin only to the target class.
More recently, OS-SCL~\cite{huang2025noise} has been proposed, which combines the mixup-based classification task with a supervised contrastive loss to further improve feature discrimination.

However, these methods have two limitations.
First, they supervise a mixed sample with the label of the first sample irrespective of the mixing ratio, so the supervision is mismatched when the first sample is only a small portion of the mixture.
Second, the angular margin is fixed irrespective of the mixing ratio, so that all mixed samples are penalized equally regardless of how much they are perturbed.

To address these issues, we reinterpret the mixup of normal samples as a perturbation of the original sample and propose PerMix, a pretext task that predicts the degree of perturbation from the original class.
To this end, we reformulate the logit of the angular margin loss so that the margin penalty is scaled according to the perturbation ratio, applying a smaller margin penalty to samples that are more perturbed. 
We further modify the supervised contrastive loss to form positive pairs based on the dominant class of each mixed sample, which resolves the mismatch between mixed samples and their assigned labels.
Since our method modifies only the training objective, it applies to both lightweight encoders~\cite{huang2025noise} and large audio models~\cite{jiang2024anopatch}.
Experiments on the DCASE 2020 Task 2 benchmark show that PerMix outperforms existing mixup-based methods with both backbones, achieving state-of-the-art results.


\section{Proposed Method}

Let $\mathcal{B}=\{1,\ldots, B\}$ be the index set of a mini-batch of normal samples, where the $i$-th sample $\mathbf{x}_i$ has a class label $y_i\in\{1,\ldots, K\}$ and $K$ is the number of classes.
An encoder $f(\cdot)$ maps $\mathbf{x}_i$ to an $\ell_2$-normalized embedding $\mathbf{z}_i=f(\mathbf{x}_i)\in\mathbb{R}^d$.
Let $\mathbf{w}_k\in\mathbb{R}^d$ denote the weight vector of class $k$. 
Then the angle between $\mathbf{z}_i$ and $\mathbf{w}_k$ is given by $\theta_{i,k}=\arccos\big(\mathbf{w}_k^\top\mathbf{z}_i/\|\mathbf{w}_k\|\big)$.

We synthesize perturbed samples by mixing each normal sample with other samples in the mini-batch.
For each anchor $\mathbf{x}_i$, we randomly select $R$ perturbing samples from the mini-batch and denote their indices by $\mathcal{J}_i\subset\mathcal{B}\setminus\{i\}$.
We first sample $R+1$ mixing coefficients from a symmetric Dirichlet distribution $\mathrm{Dir}(a)$ with $a=0.5$. 
Then we assign the largest coefficient to the anchor as $\alpha_{i, i}$, which ensures $\alpha_{i, i}\ge\alpha_{i,j}$ for all $j\in\mathcal{J}_i$, and the remaining $R$ coefficients to the perturbing samples as $\{\alpha_{i,j}\}_{j\in\mathcal{J}_i}$ in random order.
The resulting mixed sample is
\begin{equation}
\tilde{\mathbf{x}}_i=\alpha_{i,i}\mathbf{x}_i+\sum_{j\in\mathcal{J}_i}\alpha_{i,j}\mathbf{x}_j ,
\end{equation}
and its embedding and angles are denoted by $\tilde{\mathbf{z}}_i=f(\tilde{\mathbf{x}}_i)$ and $\tilde{\theta}_{i,k}$, respectively.
Since the anchor has the largest mixing coefficient, we regard $\tilde{\mathbf{x}}_i$ as a perturbed sample of its anchor class $y_i$, referred to as the \emph{dominant class}.
The ratio of the dominant class in $\tilde{\mathbf{x}}_i$ is
\begin{equation}
\lambda_i=\sum_{j\in\mathcal{J}_i\cup\{i\},\;y_j=y_i}\alpha_{i,j},
\end{equation}
which includes the coefficients of the perturbing samples of the same class as the anchor since mixing normal samples of the same class preserves the characteristics of that class.
The remaining ratio $1-\lambda_i$ comes from the other classes and is referred to as the \emph{perturbation ratio}.

\subsection{Perturbation Ratio Prediction}

Instead of predicting the mixing ratios of the classes in $\tilde{\mathbf{x}}_i$, we train the model to predict how much the dominant class is perturbed.
We reformulate the ArcFace~\cite{deng2019arcface} logit so that the margin is applied only to the dominant class and scaled by the ratio $\lambda_i$.
The logit of class $k$ for $\tilde{\mathbf{x}}_i$ is defined as
\begin{equation}
\label{eq:logit}
\ell_{i,k}=
\begin{cases}
s\big[\lambda_i\cos(\tilde{\theta}_{i,k}+m)+(1-\lambda_i)\cos\tilde{\theta}_{i,k}\big], & k=y_i,\\
s\cos\tilde{\theta}_{i,k}, & k\ne y_i,
\end{cases}
\end{equation}
where $s$ is a scale parameter, $m$ is an angular margin, and $\tilde{\theta}_{i,k}$ is the angle between the embedding $\tilde{\mathbf{z}}_i$ extracted from $\tilde{\mathbf{x}}_i$ and the weight vector $\mathbf{w}_k$.

The logit of the dominant class in \eqref{eq:logit} can be rewritten as a single cosine with an effective margin $m_i$ and an effective scale factor $\gamma_i$,
\begin{equation}
\ell_{i,y_i}=s\,\gamma_i\cos\big(\tilde{\theta}_{i,y_i}+m_i\big),
\end{equation}
where
\begin{equation}
\label{eq:eff_margin}
\begin{aligned}
m_i&=\tan^{-1}\!\Big(\frac{\sin m}{\cos m+(1-\lambda_i)/\lambda_i}\Big),\\
\gamma_i&=\sqrt{1-2\lambda_i(1-\lambda_i)(1-\cos m)}.
\end{aligned}
\end{equation}
When $\lambda_i=1$, the mixture consists only of samples of class $y_i$, and \eqref{eq:logit} reduces to the ArcFace logit.
As $\lambda_i$ decreases, both $m_i$ and $\gamma_i$ decrease, reducing the pull toward the dominant-class center.

The perturbation ratio prediction loss is defined as the binary cross-entropy between the dominant-class ratio $\lambda_i$ and the predicted probability of the dominant class $p_i=e^{\ell_{i,y_i}}/\sum_{k=1}^{K}e^{\ell_{i,k}}$:
\begin{equation}
\mathcal{L}_{\text{per}}=-\frac{1}{B}\sum_{i\in\mathcal{B}}
\big[\lambda_i\log p_i+(1-\lambda_i)\log(1-p_i)\big].
\end{equation}

\subsection{Dominance-based Contrastive Learning}
Supervised contrastive learning~\cite{khosla2020supervised} pulls together embeddings of the same class and pushes apart those of different classes.
In contrast to OS-SCL~\cite{huang2025noise}, which forms positive pairs based on the label of the first sample regardless of the mixing ratio, we assign the dominant class $y_i$ as the pseudo-label of $\tilde{\mathbf{x}}_i$ and form positive pairs among mixed samples sharing the same dominant class, i.e., $\mathcal{P}(i)=\{j\in\mathcal{B}\setminus\{i\}\mid y_j=y_i\}$.
The dominance-based contrastive loss is
\begin{equation}
\mathcal{L}_{\text{dom}}=-\frac{1}{B}\sum_{i\in\mathcal{B}}\frac{1}{|\mathcal{P}(i)|}
\sum_{j\in\mathcal{P}(i)}\log
\frac{e^{\tilde{\mathbf{z}}_i^\top\tilde{\mathbf{z}}_j/\tau}}
{\sum_{a\in\mathcal{B}\setminus\{i\}}e^{\tilde{\mathbf{z}}_i^\top\tilde{\mathbf{z}}_a/\tau}},
\end{equation}
where $|\cdot|$ denotes the cardinality of a set, and $\tau$ is a temperature parameter.

\subsection{Training and Inference}

PerMix is trained by jointly minimizing the two losses:
\begin{equation}
\mathcal{L}_{\text{PerMix}}=\mathcal{L}_{\text{per}}+\mathcal{L}_{\text{dom}}.
\end{equation}
At inference, the anomaly score of a test sample is the cosine distance to its nearest neighbor in the memory bank $\mathcal{M}=\{\mathbf{z}_n\}$ of training embeddings:
\begin{equation}
\mathrm{Score}(\mathbf{x})=\min_{\mathbf{z}_n\in\mathcal{M}}
\big(1-f(\mathbf{x})^\top\mathbf{z}_n\big).
\end{equation}

\begin{table*}[ht!]
\centering
\caption{Performance comparison of existing ASD models on the DCASE 2020 development set in terms of AUC (\%) and pAUC (\%) for each machine type. The best and second-best results are highlighted in \textbf{bold} and \underline{underlined}, respectively.}
\label{table:dcase2020_dev_performance}
\resizebox{\textwidth}{!}{
\begin{tabular}{lcccccccccccccc}
\toprule
 & \multicolumn{2}{c}{ToyCar} 
 & \multicolumn{2}{c}{ToyConveyor} 
 & \multicolumn{2}{c}{Fan} 
 & \multicolumn{2}{c}{Pump} 
 & \multicolumn{2}{c}{Slider} 
 & \multicolumn{2}{c}{Valve} 
 & \multicolumn{2}{c}{Average} \\
\cmidrule(lr){2-3} \cmidrule(lr){4-5} \cmidrule(lr){6-7} \cmidrule(lr){8-9} \cmidrule(lr){10-11} \cmidrule(lr){12-13} \cmidrule(lr){14-15}
Method
& AUC & pAUC 
& AUC & pAUC 
& AUC & pAUC 
& AUC & pAUC 
& AUC & pAUC 
& AUC & pAUC 
& AUC & pAUC \\ \midrule

STgram-MFN (ArcFace)~\cite{liu2022anomalous} & 94.44 & 87.68 & 74.57 & 63.60 & 94.04 & 88.97 & 91.94 & 81.75 & 99.55 & 97.61 & 99.64 & 98.44 & 92.36 & 86.34 \\
CLP-SCF~\cite{guan2023anomalous} & 95.85 & 90.19 & 75.21 & 62.79 & 96.98 & 93.23 & 94.97 & 87.39 & 99.57 & 97.73 & 99.89 & 99.51 & 93.75 & 88.48 \\
TASTgram-MFN (NAMix)~\cite{choi2024noisy} & 96.76 & 90.11 & 77.90 & 67.15 & \underline{98.32} & \textbf{95.34} & 95.44 & 85.99 & 99.53 & 97.50 & \underline{99.95} & \underline{99.74} & 94.65 & 89.31 \\
TFSTgram (OS-SCL)~\cite{huang2025noise} & 96.55 & 91.17 & 83.56 & 69.87 & \textbf{98.41} & \underline{94.50} & \underline{96.25} & 88.74 & \underline{99.63} & \underline{98.04} & 99.85 & 99.23 & 95.71 & 90.23 \\
\midrule
TFSTgram-MFN (PerMix) & \underline{98.65} & \underline{94.88} & \underline{85.04} & \underline{72.99} & 97.86 & 93.31 & 94.66 & \underline{88.90} & 99.58 & 97.80 & 99.57 & 98.01 & \underline{95.89} & \underline{90.98} \\
BEATs (PerMix) & \textbf{99.54} & \textbf{97.64} & \textbf{88.73} & \textbf{77.78} & 94.52 & 88.86 & \textbf{96.86} & \textbf{91.52} & \textbf{99.75} & \textbf{98.69} & \textbf{99.99} & \textbf{99.96} & \textbf{96.56} & \textbf{92.41} \\
\bottomrule
\end{tabular}}
\end{table*}

\section{Experimental Results}

\subsection{Experimental Setup}

\noindent \textbf{\textit{Datasets}}
We conduct experiments on the DCASE 2020 Task 2 dataset~\cite{koizumi2020description}, which is a widely used benchmark for ASD.
It contains recordings from six machine types, each with six or seven machine IDs.
Each combination of machine type and ID is treated as a unique class, resulting in 41 classes in total.
We also evaluate on the DCASE 2022 Task 2 dataset~\cite{dohi2022description} to assess generalization under domain shift.
It contains seven machine types, each with source and target domains that differ in operating conditions.
Each machine type is further subdivided by attributes representing its operating conditions, and each combination of machine type and attributes is treated as a unique class, resulting in 250 classes in total.

\noindent \textbf{\textit{Implementation Details}}
We apply the proposed method to two backbones: a lightweight encoder and a large audio model.
As the lightweight encoder, we adopt the MobileFaceNet-based network of OS-SCL~\cite{huang2025noise} with its feature perturbation head removed.
The model is trained for 300 epochs with a batch size of 64 and a learning rate of $1\times10^{-4}$, using an exponential moving average of the model weights~\cite{cai2021exponential}.
As the large audio model, we fine-tune BEATs$_{\text{iter3}}$~\cite{chen2022beats} with the same architecture as AnoPatch~\cite{jiang2024anopatch} for 100 epochs with a batch size of 32 and a learning rate of $1\times10^{-5}$.
Both backbones are trained with the AdamW optimizer.
The number of perturbing samples $R$ is sampled uniformly from $\{1,2,3\}$ for each anchor.
The loss-related hyperparameters follow OS-SCL~\cite{huang2025noise}: a margin $m=0.4$ and a scale $s=30$ for $\mathcal{L}_{\text{per}}$, and a temperature $\tau=0.02$ for $\mathcal{L}_{\text{dom}}$.
All experiments are conducted on a single NVIDIA RTX 4090 GPU.

\noindent \textbf{\textit{Evaluation Metrics}}
The models are evaluated using the area under the receiver operating characteristic curve (AUC) and the partial AUC (pAUC). 
The pAUC specifically measures the AUC within the low false-positive-rate region $[0,~0.1]$, providing an evaluation focused on minimizing false alarms.

\subsection{Performance Comparison}

\begin{table*}[ht!]
\centering
\caption{Ablation study on the DCASE 2020 development set in terms of AUC (\%) and pAUC (\%) for each machine type. The best results are highlighted in \textbf{bold}.}
\label{table:loss_ablation_study}
\resizebox{\textwidth}{!}{
\begin{tabular}{lcccccccccccccc}
\toprule
 & \multicolumn{2}{c}{ToyCar} 
 & \multicolumn{2}{c}{ToyConveyor} 
 & \multicolumn{2}{c}{Fan} 
 & \multicolumn{2}{c}{Pump} 
 & \multicolumn{2}{c}{Slider} 
 & \multicolumn{2}{c}{Valve} 
 & \multicolumn{2}{c}{Average} \\
\cmidrule(lr){2-3} \cmidrule(lr){4-5} \cmidrule(lr){6-7} \cmidrule(lr){8-9} \cmidrule(lr){10-11} \cmidrule(lr){12-13} \cmidrule(lr){14-15}
Method
& AUC & pAUC 
& AUC & pAUC 
& AUC & pAUC 
& AUC & pAUC 
& AUC & pAUC 
& AUC & pAUC 
& AUC & pAUC \\ \midrule
\color{gray}\textit{Model: TFSTgram-MFN} \\
\quad $\mathcal{L}_{\text{ArcFace}}$ & 95.38 & 90.41 & 75.55 & 61.87 & 88.49 & 83.25 & 93.26 & 86.75 & 99.33 & 96.46 & 98.45 & 93.05 & 91.74 & 85.30\\
\quad $\mathcal{L}_{\text{ArcFace}} + \mathcal{L}_{\text{dom}}$ & 94.79 & 89.76 & 75.41 & 62.12 & 90.77 & 84.54 & 93.60 & 87.15 & 99.34 & 96.56 & 98.36 & 93.61 & 92.05 & 85.62 \\
\quad $\mathcal{L}_{\text{per}}$ & 97.30 & 91.71 & 79.69 & 67.04 & 96.93 & 89.89 & \textbf{95.34} & \textbf{89.18} & 99.51 & 97.43 & 98.50 & 93.24 & 94.54 & 88.08 \\
\rowcolor{gray!20}
\quad $\mathcal{L}_{\text{per}} + \mathcal{L}_{\text{dom}}$ & \textbf{98.65} & \textbf{94.88} & \textbf{85.04} & \textbf{72.99} & \textbf{97.86} & \textbf{93.31} & 94.66 & 88.90 & \textbf{99.58} & \textbf{97.80} & \textbf{99.57} & \textbf{98.01} & \textbf{95.89} & \textbf{90.98} \\
\midrule
\color{gray}\textit{Model: BEATs} \\
\quad $\mathcal{L}_{\text{ArcFace}}$  & 98.61 & 94.91 & 75.09 & 64.15 & 89.19 & 82.07 & 96.51 & \textbf{91.90} & 99.65 & 98.15 & 99.10 & 96.63 & 93.03 & 87.97\\
\quad $\mathcal{L}_{\text{ArcFace}} + \mathcal{L}_{\text{dom}}$ & 98.36 & 95.01 & 75.52 & 62.90 & 90.06 & 83.67 & 95.60 & 90.93 & 99.71 & 98.45 & 99.78 & 98.83 & 93.17 & 88.30 \\
\quad $\mathcal{L}_{\text{per}}$  & 98.95 & 95.73 & 79.41 & 65.72 & 94.04 & 86.75 & 95.28 & 90.08 & \textbf{99.85} & \textbf{99.22} & 99.41 & 97.55 & 94.49 & 89.18 \\
\rowcolor{gray!20}
\quad $\mathcal{L}_{\text{per}} + \mathcal{L}_{\text{dom}}$  & \textbf{99.54} & \textbf{97.64} & \textbf{88.73} & \textbf{77.78} & \textbf{94.52} & \textbf{88.86} & \textbf{96.86} & 91.52 & {99.75} & {98.69} & \textbf{99.99} & \textbf{99.96} & \textbf{96.56} & \textbf{92.41} \\
\bottomrule
\end{tabular}}
\end{table*}

\begin{table}[ht]
\centering
\caption{Comparison of training objectives on the DCASE 2020 Task 2 dataset. Results are reported as mean $\pm$ standard deviation over three random seeds.}
\label{table:pretext_task_comparison}
\resizebox{1\columnwidth}{!}{
\begin{tabular}{lcccc}
\toprule
 & \multicolumn{2}{c}{Development set} & \multicolumn{2}{c}{Evaluation set} \\
\cmidrule(lr){2-3} \cmidrule(lr){4-5}
Method & AUC & pAUC & AUC & pAUC \\
\midrule
\color{gray}\textit{Model: TFSTgram-MFN} \\
\quad ArcFace~\cite{deng2019arcface} & 91.74$\pm$0.14 & 85.30$\pm$0.02 & 92.17$\pm$0.23 & 86.25$\pm$0.44 \\
\quad NAMix~\cite{choi2024noisy} & 94.10$\pm$0.37 & 88.49$\pm$0.30 & 94.59$\pm$0.75 & 89.67$\pm$0.77 \\
\quad OS-SCL~\cite{huang2025noise} &  94.28$\pm$0.06 & 89.22$\pm$0.11 & 95.58$\pm$0.15 & 90.73$\pm$0.21 \\
\rowcolor{gray!20}
\quad PerMix & \textbf{95.89$\pm$0.07} & \textbf{90.98$\pm$0.17} & \textbf{96.45$\pm$0.16} & \textbf{91.49$\pm$0.15} \\
\midrule
\color{gray}\textit{Model: BEATs} \\
\quad ArcFace~\cite{deng2019arcface} & 93.03$\pm$0.48 & 87.97$\pm$0.62 & 95.96$\pm$0.31 & 91.39$\pm$0.29 \\
\quad NAMix~\cite{choi2024noisy} & 93.85$\pm$0.10 & 88.53$\pm$0.03 & 96.06$\pm$0.13 & 90.80$\pm$0.24 \\
\quad OS-SCL~\cite{huang2025noise} & 94.63$\pm$0.70 & 88.92$\pm$0.62 & 95.90$\pm$0.30 & 90.68$\pm$0.41\\
\rowcolor{gray!20}
\quad PerMix & \textbf{96.56$\pm$0.21} & \textbf{92.41$\pm$0.37} & \textbf{97.99$\pm$0.22} & \textbf{93.17$\pm$0.29} \\
\bottomrule
\end{tabular}}
\end{table}
\vspace{0.4cm}

\begin{table}[ht]
\centering
\caption{Comparison of training cost with the BEATs backbone. All means denote the arithmetic mean of the average AUC (\%) and pAUC (\%) over both the development and evaluation sets.}
\label{table:training_cost}
\resizebox{0.7\columnwidth}{!}{
\begin{tabular}{lccccc}
\toprule
Method & Time/epoch (s) & All means (\%) \\
\midrule
ArcFace~\cite{deng2019arcface} & 482 & 92.09 \\
NAMix~\cite{choi2024noisy} & 483 & 92.31 \\
OS-SCL~\cite{huang2025noise} & 485 & 92.53 \\
\rowcolor{gray!20}
PerMix & 498 & 95.03 \\
\bottomrule
\end{tabular}}
\vspace{-0.3cm}
\end{table}

\begin{table}[ht]
\centering
\caption{Performance comparison on the DCASE 2022 dataset. $H_{\text{mean}}$ denotes the harmonic mean of the source AUC, target AUC, and pAUC over all machine types.}
\label{table:dcase2022}
\resizebox{0.95\columnwidth}{!}{
\begin{tabular}{lccc}
\toprule
Method & Development $H_{\text{mean}}$ & Evaluation $H_{\text{mean}}$ \\
\midrule
STgram-MFN (ArcFace)~\cite{liu2022anomalous} & 63.25 & -  \\
TWFR-GMM~\cite{guan2023time} & 70.89 & -  \\
BEATs-LoRA\cite{han2025exploring} & 72.43 & 61.69 \\
\midrule
\rowcolor{gray!20}
BEATs (PerMix) & \textbf{73.86} & \textbf{68.62} \\
\bottomrule
\end{tabular}}
\vspace{-0.3cm}
\end{table}

\begin{figure}[t!]
\centering
\begin{subfigure}{0.48\columnwidth}
    \centering
    \includegraphics[width=\linewidth]{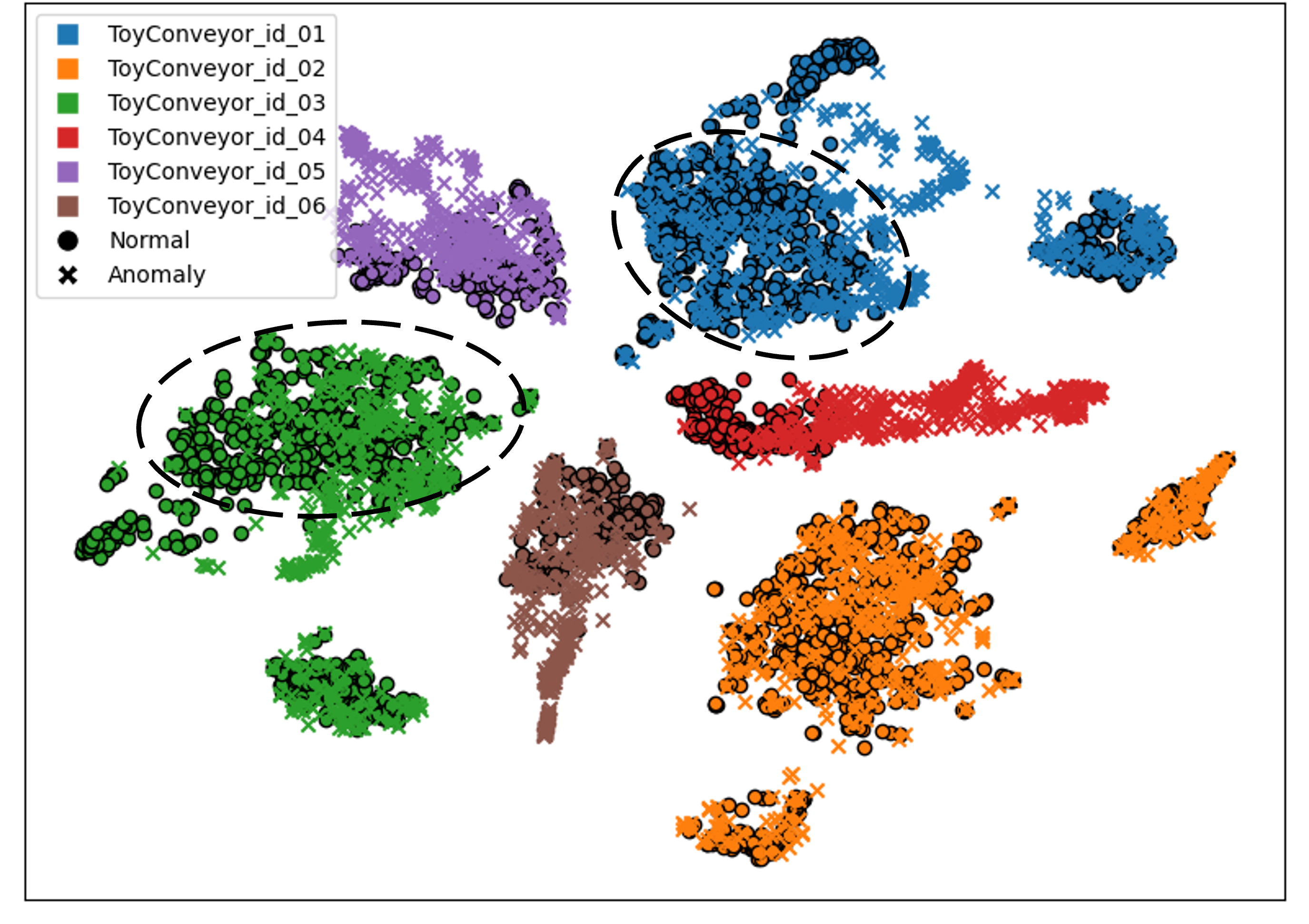}
\end{subfigure}
\hfill
\begin{subfigure}{0.48\columnwidth}
    \centering
    \includegraphics[width=\linewidth]{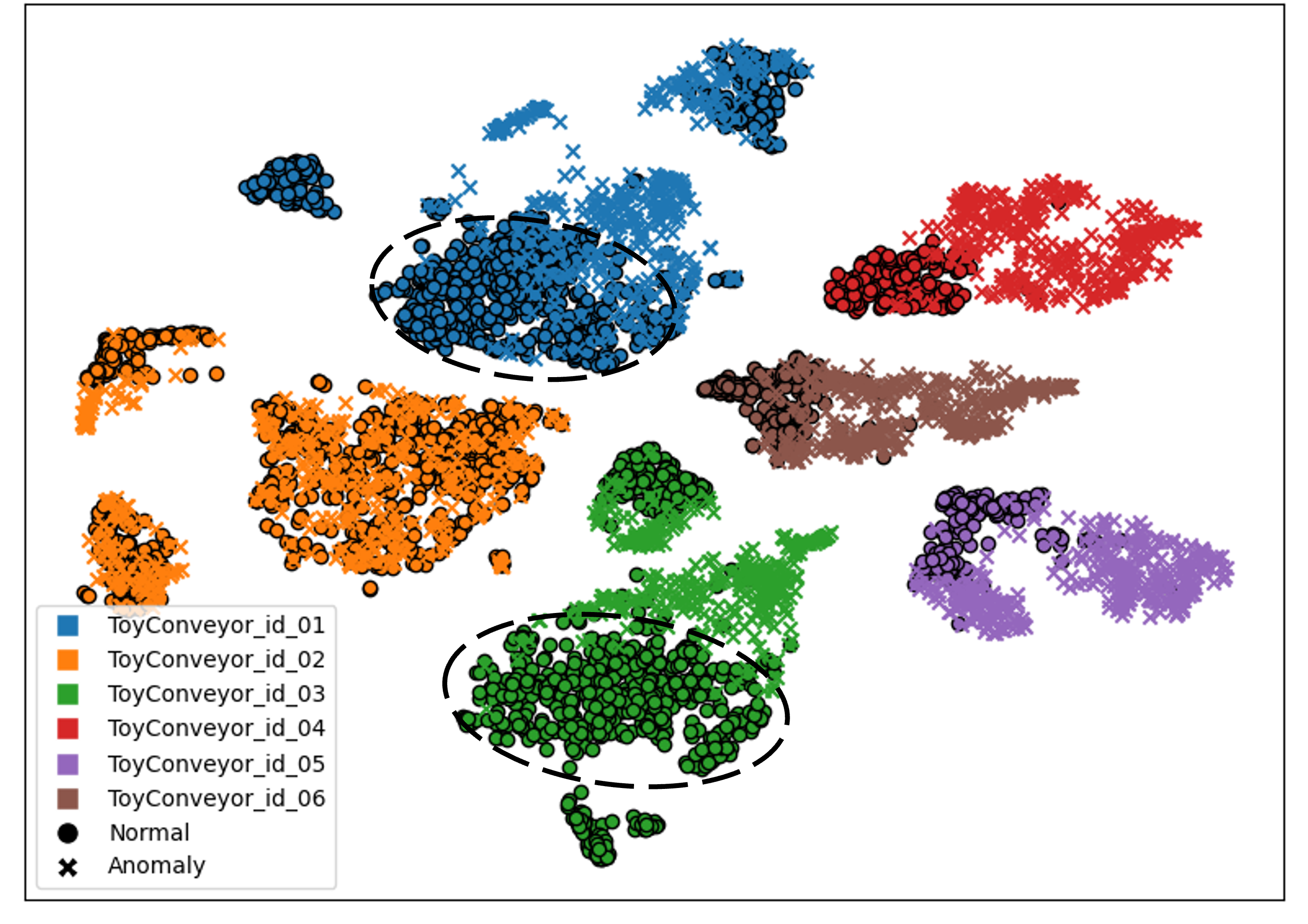}
\end{subfigure}
\includegraphics[width=1.0\columnwidth]{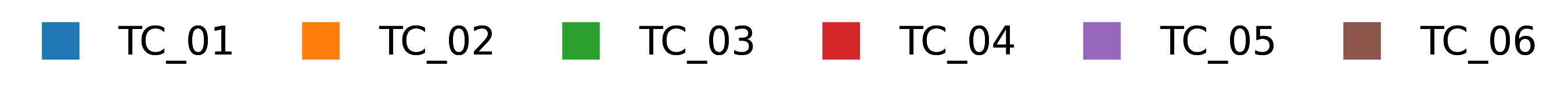}
\begin{subfigure}{0.48\columnwidth}
    \centering
    \caption{OS-SCL}
\end{subfigure}
\hfill
\begin{subfigure}{0.48\columnwidth}
    \centering
    \caption{PerMix}
\end{subfigure}
\caption{t-SNE visualization of ToyConveyor (TC) embeddings from BEATs trained with different loss functions ($\circ$: Normal, $\times$: Anomaly)}
\label{fig:tsne}
\end{figure}

Table~\ref{table:dcase2020_dev_performance} compares the proposed method with existing ASD models on the DCASE 2020 development set.
PerMix achieves the best overall performance, with an average AUC of 95.89\% using the TFSTgram-MFN backbone and 96.56\% using the fine-tuned BEATs, and outperforms the previous state-of-the-art TFSTgram (OS-SCL)~\cite{huang2025noise}.
In particular, our method shows notable improvements on ToyConveyor, a challenging machine type whose signals are dominated by background noise.
PerMix achieves a pAUC of 72.99\% with TFSTgram-MFN and 77.78\% with BEATs, the highest among the compared methods, suggesting that it learns a more compact normal data distribution.
Fig.~\ref{fig:tsne} visualizes the t-SNE projections of the ToyConveyor embeddings of BEATs trained with OS-SCL and PerMix.
OS-SCL maps several anomalous samples of ToyConveyor 01 and 03 to normal clusters (dashed circles in Fig.~\ref{fig:tsne}(a)), while PerMix keeps the normal samples in compact clusters and separates the anomalous samples from them (Fig.~\ref{fig:tsne}(b)).

To investigate the contribution of each loss, Table~\ref{table:loss_ablation_study} compares ArcFace and $\mathcal{L}_{\text{per}}$, each with and without $\mathcal{L}_{\text{dom}}$.
Replacing ArcFace with $\mathcal{L}_{\text{per}}$ substantially improves the average AUC and pAUC with both backbones, suggesting that predicting the perturbation ratio increases sensitivity to subtle anomalous signals.
Adding $\mathcal{L}_{\text{dom}}$ to ArcFace barely changes the average performance.
This is likely because ArcFace already compacts each class distribution sufficiently, so a contrastive loss that further pulls together same-class samples provides little additional supervision.
In contrast, adding $\mathcal{L}_{\text{dom}}$ to $\mathcal{L}_{\text{per}}$ improves both the average AUC and pAUC, yielding the best performance with both backbones.
The two losses play complementary roles: $\mathcal{L}_{\text{per}}$ pushes mixed samples away from the class center based on their perturbation ratio but does not pull normal samples together, while $\mathcal{L}_{\text{dom}}$ fills this gap by pulling together samples of the same dominant class, keeping the normal distribution compact.

Table~\ref{table:pretext_task_comparison} compares different training objectives on DCASE 2020 Task 2, with the backbone, training configuration, and KNN scoring kept the same.
PerMix combines perturbation ratio prediction and dominant-class contrastive learning, and improves the development set AUC over OS-SCL by 1.61 and 1.93 percentage points (pp) with TFSTgram-MFN and BEATs, respectively.
The improvement in pAUC is larger (3.49 pp with BEATs), indicating that PerMix yields fewer normal samples with high anomaly scores, consistent with the compact normal clusters in Fig.~\ref{fig:tsne}(b).
On the evaluation set with BEATs, NAMix and OS-SCL perform similarly to ArcFace, whereas PerMix improves the evaluation set AUC by 2.09 pp over OS-SCL, demonstrating the effectiveness of predicting the perturbation ratio as a pretext task.

Generating perturbed samples by mixing multiple samples can increase the training cost. 
Table~\ref{table:training_cost} compares the pretext tasks on the same BEATs backbone in terms of training time per epoch and overall mean performance.
ArcFace, NAMix, and OS-SCL show comparable training costs and similar performance of around 92\%.
PerMix requires only about 3.3\% more training time than ArcFace, yet achieves 95.03\%, 2.5 pp higher than OS-SCL.

To verify the generalization of our proposed method, we extend our experiments to the DCASE 2022 Task 2 dataset~\cite{dohi2022description}.
Table~\ref{table:dcase2022} shows the harmonic mean on the development and evaluation sets across different models.
We compare with three representative approaches: STgram-MFN (ArcFace)~\cite{liu2022anomalous}, a lightweight encoder; TWFR-GMM~\cite{guan2023time}, a non-deep-learning method that models frequency-axis statistics with GMMs; and BEATs-LoRA~\cite{han2025exploring}, a fine-tuning method for large audio models.
With the same hyperparameters as in DCASE 2020, BEATs (PerMix) achieves the highest harmonic mean on both sets.

\section{Conclusion}

In this paper, we propose PerMix, an anomalous sound detection framework that predicts the perturbation ratio of the dominant class in mixed normal samples as a pretext task.
The perturbation ratio prediction loss adjusts the angular margin according to the dominant-class ratio so that perturbed samples are not compressed toward the normal class center.
The dominance-based contrastive loss also resolves the supervision mismatch for mixed samples while pulling normal samples closer together.
PerMix outperforms existing mixup-based methods and achieves state-of-the-art performance on the DCASE 2020 Task 2 dataset with both lightweight encoder and a large audio model backbones.



\clearpage

\section{Acknowledgements}

This work was supported by the National Research Foundation of Korea (NRF) grant (No. RS-2024-00337945), the STEAM research grant (No. RS-2024-00464269) funded by the Ministry of Science and ICT of Korea government (MSIT), the BK21 FOUR program through the NRF grant funded by the Ministry of Education of Korea government (MOE), Artificial intelligence industrial convergence cluster development project funded by the Ministry of Science and ICT (MSIT, Korea)\&Gwangju Metropolitan City, and Industrial Technology Innovation R\&D program of MOTIE/KEIT (No. RS-2025-24533624).

\bibliography{ref}
\bibliographystyle{IEEEtran}





\end{document}